%% file: main.tex
\documentclass[preprint,12pt,authoryear,nopreprintline]{elsarticle}
\usepackage{lmodern}
\usepackage[margin=1in]{geometry}

\usepackage{graphicx}
\usepackage{amsmath,amssymb,amsfonts}
\DeclareMathOperator*{\argmax}{arg\,max}
\usepackage{booktabs}
\usepackage{xcolor}
\usepackage[hidelinks,hypertexnames=false]{hyperref}
\hypersetup{pdftitle={A Computational Model of Teacher-Student Co-Adaptation},pdfauthor={Francesco Balzan; Pedro P. Santos; Maurizio Gabbrielli; Mahault Albarracin; Manuel Lopes},pdfsubject={Author-prepared preprint}}
\usepackage{natbib}
\usepackage{epigraph}
\usepackage{float}
\usepackage{caption}
\usepackage{subcaption}
\usepackage{multirow}

\setcitestyle{authoryear,open={(},close={)}}

\journal{}
\date{}

\begin{document}

\begin{frontmatter}

\title{A Computational Model of Teacher-Student\\Co-Adaptation\tnoteref{preprint}}
\tnotetext[preprint]{Author-prepared preprint. Not the version of record.}
\author[unibo,pisa]{Francesco Balzan\corref{corresponding}}
\ead{francesco.balzan3@unibo.it}
\author[ist,inesc]{Pedro P. Santos}
\author[unibo]{Maurizio Gabbrielli}
\author[verses]{Mahault Albarracin}
\author[ist,inesc]{Manuel Lopes}
\address[unibo]{Department of Computer Science, University of Bologna, Italy}
\address[pisa]{Department of Computer Science, University of Pisa, Italy}
\address[ist]{Instituto Superior T\'ecnico, University of Lisbon, Portugal}
\address[inesc]{INESC-ID, Lisbon, Portugal}
\address[verses]{VERSES AI, Los Angeles, USA}
\cortext[corresponding]{Corresponding author.}

\begin{abstract}\small
Effective teaching involves mutual adaptation: teachers adjust instruction based on student responses, while students actively seek information to resolve uncertainty. Yet many computational models of teaching treat instruction as a one-way process. This paper presents a Bayesian computational framework for studying teacher-student co-adaptation under partial observability in a stylized concept-learning task. Students are organized in heterogeneous groups that differ in which features of a target concept they can observe, and the teacher infers these constraints from limited feedback. We compare five interaction modes (active teaching, active learning, turn-based interaction, adaptive teaching and full co-adaptation) and three teacher-led controls across 100 independent matched simulation runs. Learner-initiated queries resolve information gaps left by the baseline teachers' stopping rules, while teacher-led continuation also achieves complete identification in the observed episodes. Bayesian belief tracking reduces the mean interaction steps required by full co-adaptation as teacher experience accumulates. At experience level 40, full co-adaptation requires fewer interaction steps on average than active learning, turn-based interaction and all three teacher-led controls, with complete observed coverage shared by these policies. Analyses by group and student type, together with sensitivity analyses, clarify the conditions and distributional limits of these results. The framework is interpretable and modular. Its findings provide a computational proof of concept for investigating interaction structure in heterogeneous groups, rather than direct evidence about classroom practice.
\end{abstract}

 
\end{frontmatter}

\section{Introduction}\label{sec1}
 
Effective teaching is rarely a one-way process. In most educational settings, teachers adjust their instruction based on what they observe about students, while students actively contribute to the interaction by asking questions, signaling confusion, or providing feedback. This mutual adaptation between teacher and learners has been extensively studied in educational research, where sociocultural and constructivist perspectives emphasize that learning emerges through reciprocal interaction rather than passive reception \citep{Vygotsky1978-um, Bruner1996-ys, Sawyer2014-ky}. Understanding how the \emph{structure} of this interaction---who initiates, who adapts, and what information flows between participants---affects learning outcomes is a central question in both educational theory and the design of educational technologies.
 
Despite this recognition, many computational models of teaching treat instruction as a unidirectional process. In the machine teaching literature, for example, a teacher agent selects examples to guide a learner toward a target concept, but the learner has no mechanism to influence the teacher's behavior \citep{Zhu2015-js, Goldman1988-cw}. In active learning, by contrast, the learner drives the process by selecting queries, but the teacher responds passively \citep{Lewis1994-be, Sugiyama2006-ih}. Even recent Bayesian models of pedagogical reasoning, which capture mutual belief updating between teacher and learner, typically consider dyadic interaction without modeling group heterogeneity or persistent observability constraints \citep{Chen2024-sy, Shafto2014-tl}. As a result, the computational study of teaching has not yet fully examined how bidirectional interaction interacts with learner diversity under partial observability---a combination of factors that is central to educational practice but not fully addressed by these models.
 
This paper addresses that gap. We introduce a Bayesian computational framework for studying teacher--student co-adaptation under partial observability, in which a teacher interacts with groups of students who differ in what features of a target concept they can observe. The teacher must infer these observability constraints from limited feedback and adapt its instructional strategy accordingly. Students, in turn, can contribute by asking questions about features that the teacher may have overlooked. The framework is designed to be transparent and interpretable: all model parameters and decision rules are explicit, and the interaction structure is fully specified. Our specific objective is to develop and evaluate this framework to determine how the structure of teacher--student interaction---specifically, the presence or absence of learner-initiated queries and teacher belief tracking---affects task coverage and instructional efficiency in heterogeneous groups.

In this framework, \emph{co-adaptation} refers to coupled teacher and student belief updating through interaction: the teacher updates beliefs about group-specific feature observability from student feedback, while students update beliefs about the target concept and select queries according to their remaining uncertainty. Teacher observability beliefs accumulate across target concepts, whereas student concept beliefs reset for each target. The \emph{full co-adaptation} mode combines adaptive teaching with learner-initiated queries; adaptation concerns these evolving beliefs and the resulting action choices, while the underlying decision rules remain fixed.
 
We evaluate the framework using a stylized concept-learning task: a collaborative identification game in which the teacher and students work together to identify a target concept defined by binary features. This task environment is deliberately simplified. It does not attempt to reproduce the complexity of real classroom interaction, including natural language, affect, cultural context, or open-ended reasoning. Nor does it model the social, material, or cultural dimensions of education, which are central to sociocultural theories of learning but are deliberately abstracted away here to isolate the informational structure of teacher--student interaction. The value of this setting lies in its tractability and interpretability: it allows us to study, in a controlled and fully transparent environment, the computational consequences of varying the structure of interaction when learners are heterogeneous and the teacher operates under uncertainty.
 
Within this environment, we compare five interaction modes that differ systematically in who initiates communication and whether the teacher adapts to group-specific constraints:
\begin{enumerate}
    \item \emph{Active learning}---students ask questions; the teacher responds passively.
    \item \emph{Active teaching}---the teacher selects maximally informative features; students receive passively.
    \item \emph{Turn-based interaction}---teacher and students alternate initiative.
    \item \emph{Adaptive teaching}---the teacher tracks group-specific observability using Bayesian belief updating (Beta-distributed priors with Thompson sampling) and adapts feature selection accordingly.
    \item \emph{Full co-adaptation}---adaptive teaching combined with student-initiated queries.
\end{enumerate}
 
This five-mode decomposition functions as a systematic comparison: by adding or removing specific interaction components (learner queries, teacher adaptation, belief tracking), we can compare the contribution of each mechanism under explicit selection and stopping rules. Teacher-led continuation controls additionally examine the effect of early termination.
 
The study is guided by the following research questions:
 
\begin{description}
    \item[\textbf{RQ1.}] Within the modeled environment, how do learner-initiated queries and teacher stopping rules affect universal task completion when student groups differ in feature observability?
    \item[\textbf{RQ2.}] How does Bayesian belief tracking over group-specific observability affect average identification, universal completion and instructional efficiency as experience with groups accumulates?
    \item[\textbf{RQ3.}] How does the combination of adaptive teaching and learner-initiated querying compare with the other interaction modes on task coverage and efficiency?
\end{description}
 
\vspace{0.3em}
\noindent
Our main contributions are as follows:
\begin{itemize}
    \item We introduce an interpretable Bayesian computational framework for studying co-adaptive teacher--student interaction under partial observability, with explicit treatment of group-level heterogeneity. The framework extends recent models of adaptive teaching \citep{Chen2024-sy} by incorporating one-to-many interaction, group-level belief tracking via Thompson sampling, and a modular strategy comparison.
    \item We evaluate the framework through 100 independent matched simulation runs in a stylized concept-learning task with heterogeneous student groups (3 groups $\times$ 30 students, 3 student types, 15 features, ${\sim}145$ learnable concepts), with three additional teacher-led controls.
    \item We show that, within the modeled environment, learner queries overcome incomplete coverage under the baseline teacher stopping rules, Bayesian belief tracking improves full-co-adaptation efficiency with accumulated experience, and teacher-led continuation also achieves universal completion.
    \item We examine these findings across student groups and variations in observability severity, and we discuss their implications as formal hypotheses relevant to the study of interaction structure in educational settings.
\end{itemize}
 
We emphasize that the present study is a formal proof of concept. Its findings are properties of the modeled environment and do not constitute direct evidence about real classroom practice. The educational motivation for the study---the recognition that effective pedagogy involves mutual adaptation and must address learner diversity---guides the design of the framework, but the results should be interpreted within the boundaries of the simulation. Broader implications for educational research and technology are discussed as directions for future empirical investigation.
 
The remainder of the paper is organized as follows. Section~\ref{sec2} reviews computational approaches to teacher--student interaction. Section~\ref{sec:model} presents the computational model, the experimental scenario, the interaction modes, and the simulation results. Section~\ref{sec:discussion} discusses implications and connections to broader research. Section~\ref{sec:limitations} addresses limitations and future directions. Section~\ref{sec:conclusion} concludes.

\section{Computational approaches to T-SI}
\label{sec2}

Computational models have been extensively employed to investigate and understand complex systems across various disciplines, from meteorology to cognitive sciences. The epistemic strength of computational modeling resides in its capacity to operationalize a theory as a functioning system, enabling researchers to validate and test their scientific hypotheses through the creation of synthetic data and its comparison with real-world observations. This approach can be viewed as a reverse engineering process, where researchers begin with observed phenomena and strive to comprehend the causal processes that give rise to them by reconstructing these processes using computational models. However, the computational approach is not monolithic; indeed, the design of models depends on the researcher’s epistemic goals \citep{Wilson2019-vt, Kording2020-uo}.

Importantly, to enable the bidirectional transfer of insights between human and machine learning --- reviving Minsky’s interdisciplinary AI program --- computational models must balance realism and abstraction. This requires adherence to principles like \textit{psychological and neural plausibility} (aligning agent behaviors with human cognitive/neural processes); \textit{transparency} (ensuring model assumptions and parameters are interpretable to non-computational experts) and \textit{ecological validity} (simulating environments that reflect real-world dynamics). These principles ensure that models are not just mathematical abstractions but tools for reverse-engineering human behaviours (via explanatory/generative simulations). 

In the field of educational sciences, the application of the computational approach encountered many controversies (refer to \citep{Houde2022-cw} for a thorough overview). Indeed, the computational approach to human education has emphasized learners’ interactions with the environment, adhering to cognitive science methodologies, while often abstracting away the distributed, situated aspects of learning and teaching \citep{Shafto2014-tl}. Despite these limitations, computational approaches to T-SI hold significant advantages as they function as ``sandboxes'', allowing researchers to test educational hypotheses in controlled environments. These simulations enable the manipulation of variables, the exploration of hypothetical scenarios, and the generation of "inferences back to the world" that may not be feasible or ethical to examine directly in real-world educational settings. Moreover, as described below, formalized insights from computational models of T-SI can, with empirical validation, inform the design of scalable educational technologies, thereby bridging the gap between theoretical research and practical application.

For example, by the late 20th century, the rapid advances in computational models of human learning mechanisms led to applications such as cognitive tutors, which use theory-based cognitive architectures like ACT-R to enhance learning outcomes \citep{Anderson1983-fs}. This influence is bidirectional. While cognitive models shape practical tools, these educational technologies also provide valuable data that help refine theoretical learning models, creating a dynamic feedback loop between theory and application \footnote{In the ACT-R example, the computational model of human cognition and learning was used to inform the design of the first cognitive tutors whose pragmatic application in educational settings generated real data used to (in)validate and refine the hypotheses built into the ACT-R model, and so on.}.

\subsection{Non-computational insights on T-SI}

Non-computational approaches emphasize that effective T-SI is inherently co-adaptive: teachers and students co-construct learning through emotional, cultural, and cognitive negotiation. Ethnographic studies show that responsive teaching aligns pedagogical strategies with students’ lived contexts, reducing marginalization and promoting engagement \citep{Gale2002-to, Gay2010-gt, Zainullah2023-oa}. Phenomenological work emphasizes theory of mind and empathy in teaching, where students also influence instruction through feedback loops \citep{Zainullah2023-oa, Lemonie2016-vn, Mngomezulu2024-xu}. Mixed-methods studies further demonstrate that effective T-SI depends on instructional and emotional scaffolding, especially in moments where bias or misalignment could otherwise occur \citep{Pianta1995-bv, de-Ruig2024-ni, Turner2020-ds, Lorenz2021-nq}. 

These perspectives offer rich, contextualized insights, but are difficult to scale or formalize. Their narrative structure and context dependence require explicit translation choices when informing technology-enhanced learning systems or predictive computational frameworks \citep{Fenn1974-um, Tisnes2023-qm}.

\subsection{Computational Insights on T-SI}

\subsubsection{Active teaching}

Computational models of active teaching formalize how instructors strategically shape learning environments—for example, by curating examples or scaffolding tasks—to guide student progress. While these models often aim to optimize machine learning algorithms or tutor systems, some integrate insights from human education. 

Bayesian approaches have proven effective for modeling high-level cognitive dynamics in educational contexts, such as attention, retention \citep{Bertolini2023-mb}, motivation \citep{Conati2018-ry}, sampling strategies \citep{Shafto2014-tl}, and curiosity \citep{Oudeyer2016-nh}. These models leverage statistical inference and probability theory, allowing the integration of prior knowledge and historical data to predict student outcomes more accurately. Such formal frameworks support active teaching by modeling the teacher’s decision-making process in selecting pedagogically optimal samples. Using Partially Observable Markov Decision Processes (POMDPs), for example, teachers adjust strategies based on probabilistic beliefs about learner progress, maximizing expected learning gains over time \citep{Rafferty2015-fz}. %While powerful for long-term planning, POMDPs assume precise models of student cognition, which are rarely available in real classrooms. Instead, Multi-Armed Bandits (MABs) prioritize short-term adaptability, focusing on immediate feedback to optimize teaching actions (e.g., selecting exercises that resolve common errors). Hybrid models like ZPDES \citep{Clement2013-va} combine MAB efficiency with theoretical principles like Vygotsky’s “Zone of Proximal Development” (ZPD), tailoring tasks to students’ evolving skill levels. 

Other approaches integrate affective or cognitive dynamics. Models like the one proposed by \cite{Conati2009-qb} adapt teaching strategies based on inferred emotional states (e.g., frustration vs. curiosity), using probabilistic reasoning to balance engagement and difficulty. While some systems can dynamically adjust instruction by soliciting learner feedback to correct mismatches between teacher assumptions and student's needs \citep{Melo2018-qt, Guerra2021-cj}.

While these models advance formal theories of teaching, teacher-initiated formulations can leave learners as recipients of selected instruction, with limited opportunities to reshape teaching strategies through inquiry or negotiation—a core tenet of sociocultural theories \citep{Vygotsky1978-um}. For example, the hypothesis-driven teaching proposed by \cite{Shafto2014-tl} assumes teachers unilaterally select examples to steer learners toward correct answers, ignoring how student questions or misconceptions might redirect the lesson. These particular formulations do not capture the full mutual adaptation observed in human T-SI (e.g., teachers refining strategies based on student-led questions).

\subsubsection{Models of Active Learning}

Active learning—where learners guide their inquiry through questioning, exploration, or data selection—has been studied in both human education and ML. Educational active learning encompasses broader participation and engagement, whereas the computational term here denotes query selection. Computational approaches, however, diverge into two paradigms: those modeling human teacher–student interaction (T-SI), and those prioritizing algorithmic efficiency.

Early work sought to formalize active learning as observed in education. Clouse and Utgoff’s advising model \citep{Clouse1992-bg}, for instance, allowed simulated learners to request teacher guidance when uncertain, reflecting metacognitive self-regulation in humans \citep{Metcalfe2009-dn}. Curiosity-driven frameworks \citep{Sun2022-qg} emulate intrinsic motivation such as novelty-seeking, testing hypotheses about how curiosity enhances learning. These approaches align with qualitative research emphasizing the importance of student agency and active learning for education \citep{Raes2022-ri}. Reinforcement learning (RL) models further capture how learners balance exploration (seeking new information) with exploitation (leveraging known strategies), providing mechanistic accounts of human pedagogical dynamics \citep{Sutton1999-jy}. All these approaches prioritize psychological plausibility, grounding computational choices in cognitive and educational theory.

In contrast, applied active learning research primarily targets data efficiency. Its benefits are well-established across domains: in inverse reinforcement learning, active querying accelerates concept acquisition \citep{Cakmak2010-hr}; in regression, it reduces the number of examples required for accurate prediction \citep{Sugiyama2006-ih}; and in classification, uncertainty sampling and query-by-committee achieve substantial efficiency gains \citep{Lewis1994-be}. Expected-error-reduction strategies extend these improvements to sample selection \citep{Roy2001-ua}, and theoretical work confirms their efficiency guarantees \citep{Hanneke2011-md}. Yet, these models often abstract away human cognition. Robotic active learning, for example, treats exploration as raw data acquisition \citep{Taylor2021-ai}, omitting the teacher’s role in scaffolding inquiry. Similarly, curiosity-driven RL agents \citep{Pathak2017-gr} optimize for novelty without modeling the metacognitive or social dimensions of curiosity. Effective for ML, these approaches require empirical grounding before informing human T-SI. 

Importantly, many of these approaches do not aim to explain or support human learning. Instead, they often rely on simplified or naive metaphors of education—such as “students asking questions” or “curiosity-driven exploration”—primarily as inspiration to improve machine learning performance. As a result, their pedagogical realism is limited, and their insights cannot easily be transferred back to understanding teacher–student interaction.

\subsubsection{Models of Active Teaching and Learning}

More recently, multi-agent approaches have sought to combine active teaching and learning. Most, however, emphasize computational efficiency over educational realism. In multi-agent reinforcement learning (MARL), for example, agents are rewarded for social influence but fall short of capturing the negotiation that characterizes real classrooms \citep{Jaques2019-fi}. Interactive POMDPs extend this by embedding theory-of-mind reasoning through nested beliefs, yet their focus remains on robotic collaboration rather than human pedagogy \citep{Han2018-uu}. Similarly, Amir et al.’s jointly-initiated advising framework \citep{Amir_undated-js} proposes strategies such as “Ask Important–Correct Important” to balance teacher attention and learner autonomy, but its design targets autonomous systems like self-driving cars, not human education. Concepts like ``attention'' and ``autonomy'' are borrowed from T-SI research, yet abstracted away from their cognitive and sociocultural contexts.

Here too, the use of educational concepts is largely metaphorical. While terms such as “attention,” “autonomy,” or “scaffolding” are borrowed from educational theory, they are typically redefined to optimize artificial agents rather than to illuminate human pedagogy. This creates a gap between computational rigor and the socio-cognitive realities of classrooms.

A relevant computational account of teacher--student interaction is the hierarchical Bayesian model of adaptive teaching proposed by \citet{Chen2024-sy}. This framework captures co-adaptation through mutual belief updates, addressing both learners’ uncertainty about concepts and higher-order uncertainty about each other’s knowledge and intentions. It enables teachers to tailor examples to learner backgrounds while students provide strategic feedback. However, it remains limited: interactions are short-term and real-world diversity (e.g., sensory or cultural differences) is not modeled.

\bigbreak

Taken together, these literatures motivate further computational study of heterogeneous groups, information access and adaptation over repeated interactions. The flow of insights between education and machine learning depends on the assumptions and validation of each model. In the next section, we present our model, which addresses a restricted part of these questions by simulating heterogeneous groups where teacher observability beliefs and student concept beliefs are coupled over time. The framework does not implement contextual scaffolding or formalise sociocultural theory.

\section{Computational Model and Experimental Evaluation}
\label{sec:model}
 
This section presents the computational framework, the experimental scenario used to evaluate it, and the simulation results. We first describe the task environment and the assumptions about student heterogeneity and teacher observability. We then define the five interaction modes compared in the study. Finally, we report the simulation results addressing each of the three research questions stated in Section~\ref{sec1}.
 
The framework adopts a generative modeling approach guided by the following design principles:
\begin{itemize}
    \item \textit{Transparency}: all model assumptions, parameters, and decision rules are fully specified and interpretable.
    \item \textit{Modularity}: interaction components (teacher initiative, student queries, belief tracking) can be independently added or removed, enabling systematic comparison.
    \item \textit{Formal grounding}: agents update beliefs via Bayesian inference, providing a principled basis for adaptive decision-making under uncertainty.
\end{itemize}
 
We note that while the framework is motivated by the educational observation that effective teaching involves mutual adaptation, the model itself implements an information-theoretic decision process. Connections to educational theory are discussed where appropriate, but the model should be understood as a formal and computational tool, not as a direct operationalization of sociocultural theory. Table~\ref{tab:construct_mapping} makes this relationship explicit.

\begin{table}[!htbp]
\centering
\caption{Mapping between educational concepts and computational implementation. The rightmost column identifies where the analogy between theory and model is limited.}
\label{tab:construct_mapping}
\footnotesize
\setlength{\tabcolsep}{4pt}
\begin{tabular}{p{2.5cm} p{3.2cm} p{3.0cm} p{3.5cm}}
\toprule
\textbf{Educational concept} & \textbf{Computational implementation} & \textbf{What the mapping captures} & \textbf{Where the analogy breaks down} \\
\midrule
Learner diversity & Binary observability types (each type blind to 5 of 15 features) & Differential access to information & Does not capture cognitive, affective, or cultural differences \\
\addlinespace
Guided instruction / scaffolding & Information-gain-based feature selection & Teacher provides most informative cues & No dynamic difficulty adjustment; no linguistic scaffolding \\
\addlinespace
Learner agency / inquiry & Student selects a feature to ask about & Learner contributes information to the interaction & Limited to predefined feature queries; no open-ended inquiry \\
\addlinespace
Adaptive teaching & Bayesian belief tracking (Beta priors) + Thompson sampling & Teacher adjusts strategy based on accumulated evidence & Assumes rational updating; no teacher uncertainty about goals \\
\addlinespace
Zone of proximal development & Not directly implemented & Effective instruction targets what the learner can almost achieve & No ZPD mechanism in the model; inspiration only \\
\addlinespace
Equitable outcomes & Universal task completion across all groups & All students reach correct identification & Does not capture equity of access, effort, or participation \\
\bottomrule
\end{tabular}
\end{table}

\subsection{Experimental Scenario}
\label{sec:scenario}
 
We evaluate the framework using a collaborative concept-learning task inspired by the ``Guess Who?'' game. A teacher interacts with groups of students to help them identify a target concept defined by binary features. This task environment is deliberately simplified: it provides a tractable and fully interpretable setting for studying how interaction structure affects learning outcomes under heterogeneous partial observability, without attempting to model the complexity of real classroom interaction.
 
We refer the reader to Supplementary Methods S2, Section~1 for the full mathematical specification. Below, we describe the key components of the experimental scenario. Table~\ref{tab:exp_config} summarizes all parameter values used in the simulations.
 
\paragraph{Concept space}
Each simulation run generates a concept space by sampling binary feature vectors for 500 candidate concepts, where each feature is active with a specified probability (see Table~\ref{tab:exp_config}). After removing duplicate concepts and filtering for learnability (ensuring all concepts remain distinguishable from the perspective of every student type), the number of unique learnable concepts ranged from 115 to 188 across the 100 primary runs ($M = 145.45$, $SD = 13.30$). After duplicate removal, student types 1, 2 and 3 are processed sequentially: for each type, the first representative of every distinct observable projection is retained using \texttt{numpy.unique}. The type order and preceding row order can therefore affect the retained set. Subsequent removal preserves distinguishability for earlier types.
 
\paragraph{Student heterogeneity}
Each student is assigned one of three types. Each type is characterized by a set of 5 features (out of 15) that students of that type cannot observe. These ``blind'' features are randomly selected at the beginning of each simulation run. This observability constraint serves as a simple, abstract representation of heterogeneous access to information. We do not claim that it captures the full complexity of real-world perceptual, cognitive, or cultural differences; rather, it provides a minimal and tractable formalization of the idea that learners in the same group may differ in what they can extract from the same instructional content.
 
\paragraph{Group structure}
Students are organized into three groups of 30 (90 students total). Each group has a dominant student type: 90\% of students in a group are drawn from one type, with 5\% from each of the other two types. This creates groups with distinct but overlapping observability profiles. The teacher knows which group it is interacting with, but does not initially know which features are unobservable for that group.
 
\begin{figure}[!htbp]
    \centering
    \includegraphics[width=0.95\linewidth]{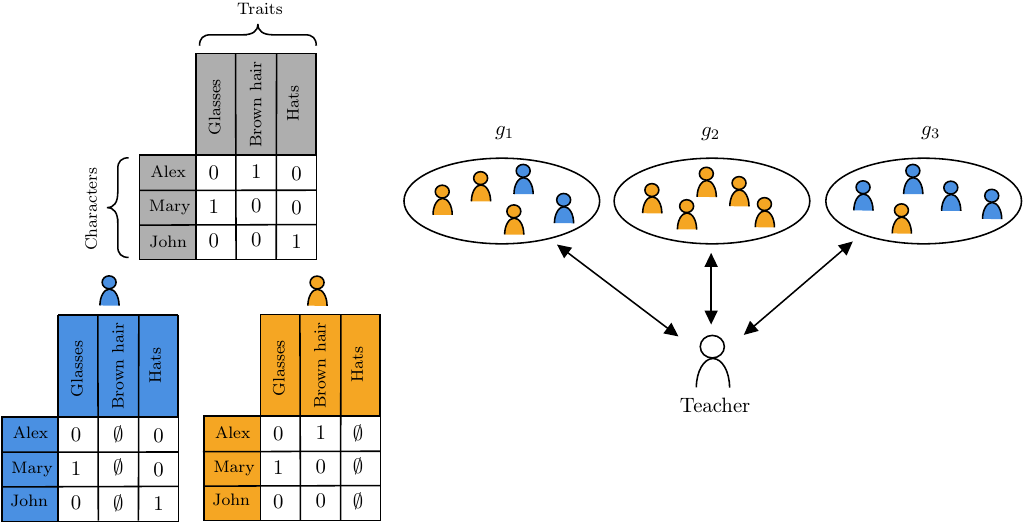}
    \caption{Illustrative schematic of the teacher--student interaction 
    setting. For visual clarity, the schematic uses a simplified 
    configuration (15 students in 3 groups of 5, two student types, 
    three concepts, three features); the actual simulations use a larger 
    setting (90 students in 3 groups of 30, three student types, 
    {\textasciitilde}150 learnable concepts, 15 features), as specified in 
    Table~\ref{tab:exp_config}. The top-left table shows the binary feature 
    values for each of the three example concepts (Alex, Mary, John). 
    The set of student types is $\{$\raisebox{-.2ex}{\includegraphics[height=2ex]{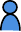}}, \raisebox{-.2ex}{\includegraphics[height=2ex]{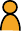}}$\}$. 
    The bottom-left tables display the concepts from the perspective 
    of each type: students of type \raisebox{-.2ex}{\includegraphics[height=2ex]{student-type-blue.pdf}} 
    cannot distinguish hair color, and students of type 
    \raisebox{-.2ex}{\includegraphics[height=2ex]{student-type-orange.pdf}} 
    cannot distinguish whether characters wear a hat. Each student 
    type is therefore unable to observe a subset of the features, 
    illustrating the partial-observability assumption central to the 
    framework.}
    \label{fig:teacher-students-interaction-illustration}
\end{figure}

\paragraph{Interaction protocol}
The teacher sequentially teaches 41 target concepts (sampled uniformly at random with replacement) to all three groups. For each target concept, the teacher interacts with each group in sequence. Within a group, the teacher and students exchange information according to the specified interaction mode until either all students have correctly identified the target concept, or the maximum number of interaction steps (15, equal to the number of features) is reached; teacher-only policies may stop earlier under their own rule. One step is one actual feature/value, initiated by the teacher or by an answered student query. The sampled observability acknowledgement is included in that exchange; switching initiative has no communication cost. The evaluator stops observation once all students identify the target, without giving hidden individual posteriors to the teacher for feature selection. In adaptive modes, the teacher's Bayesian beliefs about group-specific observability accumulate across target concepts, modeling the growth of teaching experience. Figure~\ref{fig:teacher-students-interaction-illustration} provides a schematic illustration of the interaction setting.

\paragraph{Information structure}
The distinction between what is observable and what must be inferred is central to the framework. Table~\ref{tab:info_structure} summarizes the information available to each agent.
 
\subsection{Interaction Modes}
\label{sec:modes}
 
We compare five interaction modes that differ systematically in who initiates communication and whether the teacher adapts to group-specific constraints. Table~\ref{tab:strategy_comparison} provides a summary.

\paragraph{Active learning}
In this mode, only students ask questions. At each step, a randomly selected student who has not yet identified the target concept selects the feature with the highest expected information gain and asks the teacher for its value. The teacher responds truthfully. All students in the group hear the question and the response and update their probability distributions over concepts accordingly. The interaction continues until all students have correctly identified the target concept or the interaction budget is exhausted. Students who cannot observe a given feature do not update their beliefs based on that feature's value.
 
\paragraph{Active teaching}
Here, the teacher takes initiative. At each step, the teacher selects the feature with the highest information gain (given the current probability distribution over concepts) and announces its value to all students. A randomly selected student reports back to the teacher whether they can observe the suggested feature. If the student cannot observe it, the teacher records this and excludes the feature from future suggestions within the current episode. The teacher's probability distribution is updated only when the responding student confirms observation. The interaction ends when the teacher's probability distribution is concentrated on the correct concept or no informative non-excluded feature remains. Importantly, the teacher may stop prematurely, because some students may still lack sufficient information due to unobservable features.
 
\paragraph{Turn-based interaction}
This mode interleaves teacher initiative and student queries. The teacher begins with active teaching. When the responding student reports not observing a suggested feature, initiative passes to the students for one active learning step, after which the teacher resumes. If the teacher concludes the interaction (believing all students are correct) while some students have not yet identified the target, the remaining steps proceed as active learning. The interaction ends when all students have correctly identified the target or the interaction budget is exhausted.
 
\paragraph{Adaptive teaching}
In adaptive teaching, the teacher maintains Bayesian beliefs about each group's observability. For each feature $\phi$ and group $g$, the teacher tracks a Beta distribution $\text{Beta}(\alpha_{\phi,g}, \beta_{\phi,g})$ over the probability that students in group $g$ can observe feature $\phi$. These parameters are initialized to $\alpha = \beta = 1$ (uniform prior) and updated after each group interaction based on whether the responding student reported observing each suggested feature. Before interacting with a group, the teacher samples estimated observability values $\hat{\theta}_{\phi,g} \sim \text{Beta}(\alpha_{\phi,g}, \beta_{\phi,g})$ via Thompson sampling, and selects the feature maximizing \emph{expected} information gain, weighted by $\hat{\theta}_{\phi,g}$. This introduces an exploration--exploitation tradeoff: the teacher sometimes selects features that appear suboptimal given current estimates in order to refine its beliefs. As with non-adaptive teaching, the teacher may stop the interaction before all students have identified the target.
 
\paragraph{Full co-adaptation}
This mode combines adaptive teaching with student-initiated queries, using the same interleaving logic as the turn-based mode. The teacher uses Bayesian belief tracking and Thompson sampling for feature selection, and students contribute queries when the teacher's suggestions fail. This mode represents the richest form of interaction in the framework: the teacher adapts to group-specific constraints, and students compensate for the teacher's residual uncertainty.

\bigbreak

Here, co-adaptation denotes this coupling of belief updates, not a separate policy-learning algorithm. Student concept beliefs and the teacher's group probability distribution reset for each target/group episode; the student query-selection rule remains fixed. Group-feature observability beliefs persist across targets and update after each group episode using the sampled respondents' acknowledgements to teacher-led interaction steps. Answered learner queries update the teacher's concept probability distribution but not the observability counters. A teacher-to-query transition occurs immediately within the same interaction budget.

To distinguish student querying from early teacher termination, we also compare non-adaptive and adaptive teacher-led continuation controls, which present unannounced features in ascending index order after the baseline teacher stopping condition, and exhaustive teaching, which follows that order from the beginning. They use the same budget of 15 interaction steps and no learner queries or hidden individual beliefs. Full specifications and results appear in \ref{app:r2controls}.

\subsection{Simulation Results}
\label{sec:results}
 
\begin{figure}[!htbp]
    \centering
    \includegraphics[width=\textwidth]{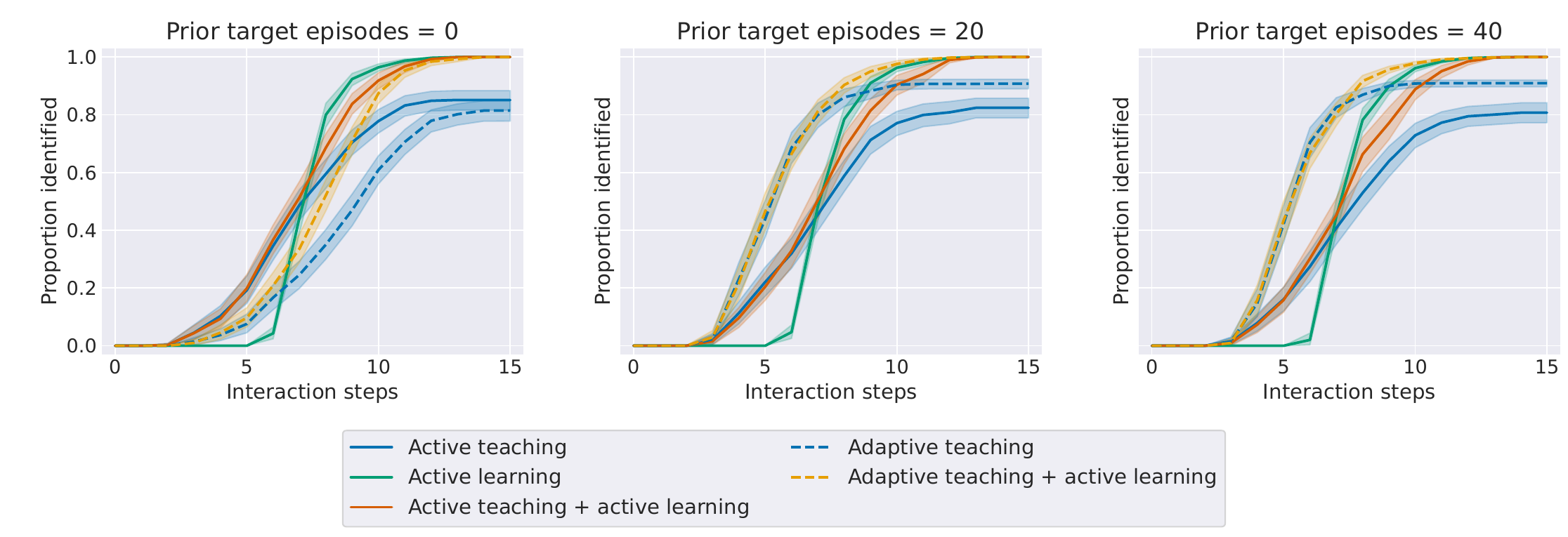}
    \caption{Mean percentage of students correctly identifying the target concept across interaction steps for each of the five interaction modes. Shaded regions show pointwise 95\% percentile intervals from 30,000 whole-run resamples, preserving the three dependent groups and matched strategies in each of 100 independent runs. For episodes ending before step 15, the final percentage is retained at subsequent plotted steps when calculating the average. Left: no prior experience (0 concepts previously taught). Middle: moderate experience (20 concepts). Right: experienced teacher (40 concepts). Dashed lines indicate adaptive modes (with Bayesian belief tracking); solid lines indicate non-adaptive modes.}
    \label{fig:exp_results_1}
\end{figure}

We conducted 100 independent simulation runs, with the same concept space, masks, groups and target sequence shared across the five interaction modes and three controls within each run. All conditions used the same parameter settings (Table~\ref{tab:exp_config}), differing only in the interaction logic. Figure~\ref{fig:exp_results_1} shows the mean percentage of students who correctly identified the target concept at each actual interaction step. Shaded regions indicate pointwise 95\% bootstrapped confidence intervals from 30,000 resamples of the 100 entire runs, retaining their groups, targets and matched strategies together. The three panels in the figure correspond to increasing levels of teacher experience: 0, 20, and 40 prior target concepts taught. Table~\ref{tab:results_summary} provides a quantitative summary. Mean trajectory coverage includes all episodes, including failures, by averaging identification over 15 plotted steps with terminal states carried forward; it does not impute a completion time to failures. Direct contrasts preserve run pairing; a summary of the uncertainty procedure appears in \ref{app:statistics}, with the full derivation in Supplementary Methods S2, Section~2.

\input{r2_primary_table.tex}

\paragraph{RQ1: Learner queries and universal task completion}
Active teaching alone reaches a mean of $80.72\%$ correct students at experience 40 and achieves universal task completion in $72/300$ group-episodes. By contrast, all three modes incorporating learner queries---active learning, turn-based, and full co-adaptation---reach $100\%$ correct and complete $300/300$ episodes (Table~\ref{tab:results_summary}). The explanation for incomplete teacher-only coverage is structural: the teacher, operating under partial observability, may conclude the interaction before all students have acquired sufficient information. Learner-initiated queries allow students who have not yet identified the target to seek the specific features they need. However, all three teacher-led controls also complete $300/300$ episodes (Table~\ref{tab:r2_controls}). Since filtering guarantees distinguishability for every type, announcing all features once is sufficient without queries.
 
\paragraph{RQ2: Belief tracking improves efficiency with experience}
Comparing non-adaptive and adaptive modes, the benefit of Bayesian belief tracking is evident in full-co-adaptation completion time as teacher experience grows. Mean interaction steps to 100\% decrease from $10.03$ ($SD = 1.84$) at experience 0 to $9.12$ ($SD = 2.13$) at experience 40---a reduction of approximately one interaction step, or 9\%. Adaptive teaching without learner queries instead increases average final identification from $81.47\%$ to $90.96\%$ while universal completion decreases from $102/300$ to $40/300$. Its successful-episode time falls from $10.18$ to $6.10$ interaction steps, but the shrinking successful subset does not establish unconditional efficiency. At experience 40, dominant-type identification is $99.01\%$ ($8064/8145$ students), versus $14.27\%$ ($122/855$) for minority types. These outcomes show why average identification, whole-group completion and communication cost must be interpreted separately; the within-group breakdown appears in \ref{app:r2subgroups}.
 
\paragraph{RQ3: Coverage and efficiency of full co-adaptation}
At experience level 40, full co-adaptation achieves $100\%$ identification and completes $300/300$ episodes, with a mean of $9.12$ interaction steps ($SD = 2.13$). Active learning achieves the same observed completion but requires $10.07$ interaction steps ($SD = 1.72$); the paired difference is $-0.94$ (95\% interval $[-1.12,-0.77]$). Turn-based interaction also completes every episode, requiring $9.72$ interaction steps; the corresponding difference is $-0.60$ ($[-0.76,-0.43]$). These time contrasts include the full common episode set because both policies complete every episode. Compared with the three teacher-led controls, full co-adaptation requires fewer interaction steps on average: the paired differences are $-2.65$ (95\% interval $[-2.94,-2.37]$) against active teaching with continuation, $-2.75$ ($[-3.02,-2.49]$) against adaptive teaching with continuation, and $-4.56$ ($[-4.85,-4.27]$) against exhaustive teaching (Table~\ref{tab:r2_controls}). All four policies complete $300/300$ episodes. Against adaptive teaching with continuation, however, the mean trajectory-coverage difference is only $0.52$ percentage points, with an interval spanning zero ($[-0.04,1.11]$). Thus full co-adaptation combines complete observed coverage with lower communication cost in these comparisons, without superior completion or superiority on every outcome. Direct contrasts are reported in \ref{app:r2contrasts}.

\paragraph{Consistency across groups}
To assess whether the results hold at the group level, we examined per-group performance at experience 40 (Table~\ref{tab:group_results}). Under full co-adaptation, all three groups reach $100\%$ correct with mean completion costs between $8.96$ and $9.21$ interaction steps, with all 100 observed episodes completed in each group. Under active teaching, final identification ranges from $75.90\%$ to $86.33\%$. These group averages can still conceal differences between student types, which are reported separately.

\begin{figure}[!htbp]
    \centering
    \includegraphics[width=\textwidth]{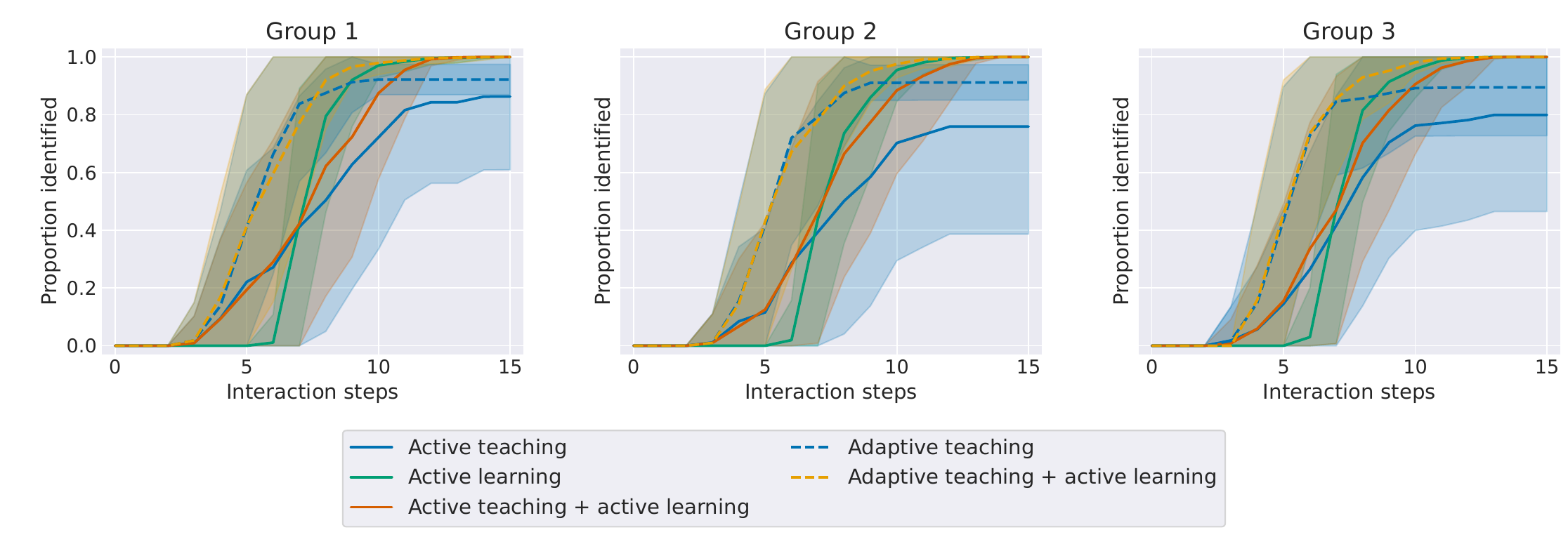}
    \caption{Per-group performance trajectories at experience level 40 (40 concepts previously taught). Each panel shows one student group. Lines show mean identification across 100 independent simulation runs; shaded regions show $\pm 1$ SD across runs and describe between-run variability rather than confidence intervals for the mean. For episodes ending before step 15, the final identification proportion is retained at subsequent plotted steps. Under full co-adaptation, each group completes all 100 observed episodes; these descriptive results do not establish equivalence between group trajectories.}
    \label{fig:group_level}
\end{figure}
 
\paragraph{Robustness to observability severity}
To assess whether the main findings depend on the specific observability configuration used in the primary analysis, we evaluated the five core strategies and three controls on 30 matched runs per condition, varying the number of unobservable features per student type: 3, 5 (original), and 7 out of 15. The five-disabled reference uses the first 30 primary runs. Table~\ref{tab:robustness} reports the five core strategies at experience 40; complete eight-policy outputs accompany the code. Active learning and full co-adaptation complete all 90 group-episodes in each condition, whereas the baseline teacher-only policies remain below universal completion. Full-co-adaptation completion times are $9.06$, $8.90$ and $8.60$ interaction steps, compared with $10.24$, $9.73$ and $9.09$ for active learning. Filtering also changes the retained concept set: mean counts are $267.70$, $143.60$ and $56.63$, with mean initial entropies $8.06$, $7.16$ and $5.82$ bits. Counts and a common exhaustive-policy reference characterise, but do not fully equalise, task difficulty. Severity effects therefore cannot be separated from task-space changes (\ref{app:r2counts}).

\begin{figure}[!htbp]
    \centering
    \includegraphics[width=\textwidth]{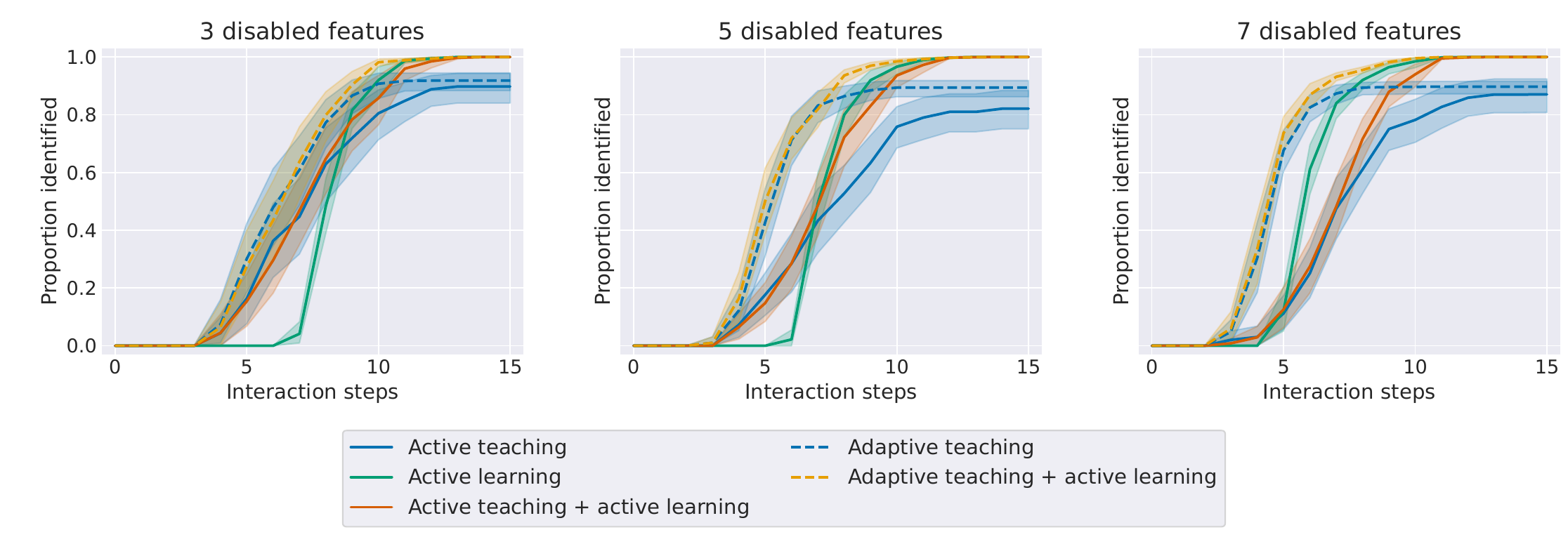}
    \caption{Robustness to observability severity at experience level 40. 
    Each panel shows results for a different number of unobservable 
    features per student type (3, 5, and 7 out of 15). Each condition uses 30 matched runs; shading gives pointwise whole-run 95\% bootstrap intervals. The baseline teacher-only policies remain below universal completion, while active learning and full co-adaptation complete all observed episodes. Concept-space filtering varies across conditions, so difficulty is not held fixed.}
    \label{fig:robustness}
\end{figure}

\begin{figure}[!htbp]
    \centering
    \includegraphics[width=\textwidth]{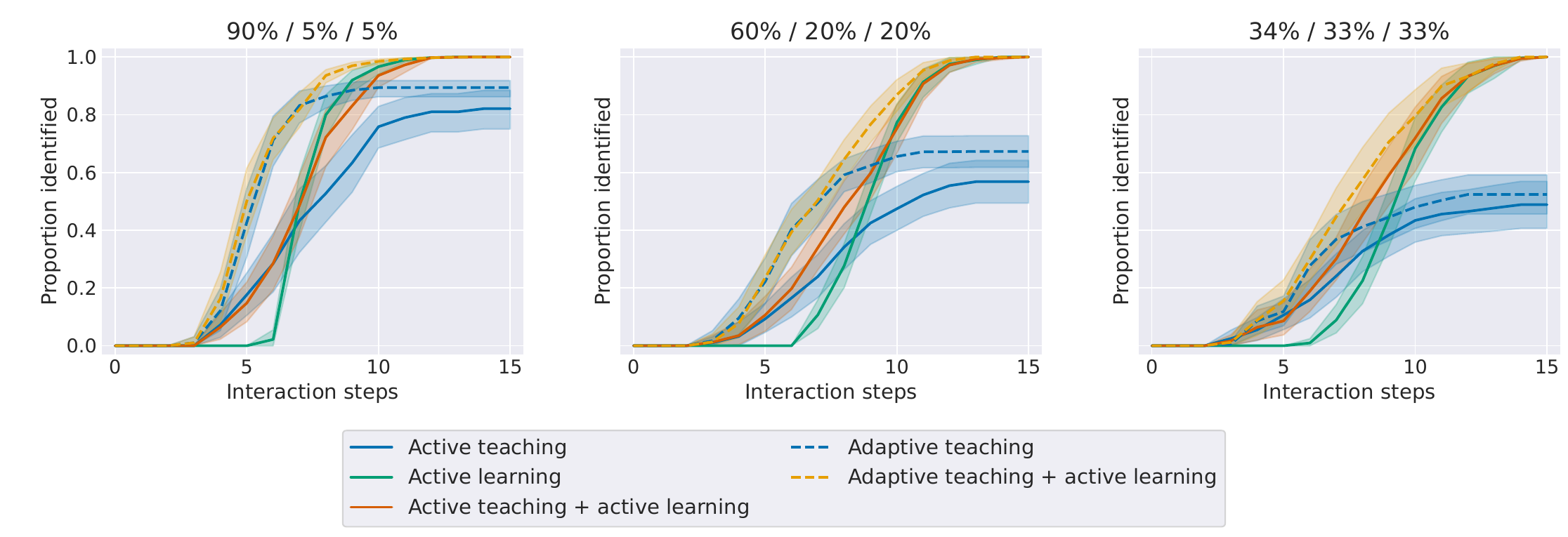}
    \caption{Robustness to group homogeneity at experience level 40. 
    Each panel shows results for a different within-group type distribution: 
    high homogeneity (90\%/5\%/5\%, matching the primary analysis), 
    moderate (60\%/20\%/20\%), and uniform (34\%/33\%/33\%). 
    Each condition uses 30 matched runs; shading gives pointwise 95\% whole-run bootstrap intervals. The five core strategies are displayed.}
    \label{fig:robustness_homogeneity}
\end{figure}

\paragraph{Robustness to group homogeneity}
We further examined whether the advantage of co-adaptation depends on the degree of within-group homogeneity. In the primary analysis, each group is dominated by a single student type (90\%/5\%/5\%). We evaluated all eight policies on 30 matched runs under moderate heterogeneity (60\%/20\%/20\%) and approximately uniform composition (34\%/33\%/33\%), using the first 30 primary runs as the high-homogeneity reference. Figure~\ref{fig:robustness_homogeneity} shows the five core strategies at experience 40. The query-based strategies retain full observed completion across these configurations; efficiency and early-trajectory differences depend on the composition. The executable supplement reports all eight policies and all-episode outcomes, rather than assuming a ranking shared by every metric.
 
\section{Discussion}
\label{sec:discussion}
 
The simulation results show that, within the stylized concept-learning task studied here, the structure of teacher--student interaction has a substantial effect on learning outcomes when students differ in what information they can access. Learner-initiated queries allow full co-adaptation to address information gaps remaining when the teacher’s selection rule stops. At experience level 40, full co-adaptation achieves universal completion in fewer interaction steps on average than the tested teacher-led continuation strategies, which also reach universal completion. Bayesian belief tracking reduces full-co-adaptation completion cost with accumulated experience, although average and universal outcomes can diverge (RQ2). Combining adaptation and queries reduces communication cost relative to the stated comparators without outperforming them on every outcome (RQ3). Group and sensitivity results are reported in Tables~\ref{tab:group_results} and \ref{tab:robustness}. In this section, we discuss what the simulation demonstrates, what it does not demonstrate, and what broader implications may be worth investigating in future work.
 
\subsection{What the Simulation Demonstrates}
 
The most direct finding is that the baseline early-stopping teacher policies under partial observability are insufficient for universal task completion. When the teacher selects features based on information gain but cannot observe which features students actually perceive, approximately $19\%$ of students remain unresolved. Non-adaptive teaching does not use group-specific information and remains below full coverage at all three experience levels. By contrast, allowing students to ask questions about features they cannot resolve ensures that all students succeed in the observed query-based episodes, because unresolved students continue to seek relevant information. The successful teacher-led controls show that questions are one route to complete coverage rather than a necessary condition.
 
Adaptive belief tracking adds a complementary benefit. By maintaining Beta-distributed estimates of each group's observability and using Thompson sampling for feature selection, the teacher progressively learns which features are likely to be effective for each group. This reduces the number of interaction steps required for universal completion from $M = 10.03$ to $M = 9.12$ interaction steps as experience accumulates---a meaningful efficiency gain in the modeled environment, corresponding to approximately one fewer interaction per group-concept episode.

The teacher-only adaptive policy also reveals an important divergence between aggregate and within-group outcomes. At experience level 40, identification reaches $99.01\%$ among dominant-type students but only $14.27\%$ among minority-type students (Table~\ref{tab:r2_subgroups}). Thus, improved average identification does not imply that all student types benefit comparably. This finding concerns the tested group-level adaptation and stopping rules; it does not establish an inherent disadvantage of teacher-led interaction, since the continuation controls also achieve complete identification.
 
The five-mode decomposition also enables an ablation-like analysis. Comparing full co-adaptation with adaptive teaching reveals the effect of querying together with continued interaction after the teacher stops. The continuation controls separate early termination from an absolute need for queries. At experience level 40, full co-adaptation requires fewer interaction steps on average than turn-based interaction. Removing both mechanisms yields the baseline active-teaching policy, which does not consistently achieve universal completion under its stopping rule. This modular structure supports controlled comparisons of specified interaction rules; it does not isolate an abstract effect of initiative independent of feature selection and stopping.
 
\subsection{What This Study Does Not Demonstrate}
 
It is important to state explicitly what the present findings do not establish.
 
First, the model does not capture the complexity of real teacher--student interaction. Real pedagogy involves natural language, affect, motivation, metacognition, cultural context, classroom management, institutional constraints, and many other dimensions that are absent from the simulated environment. The concept-learning task is deliberately simplified: students select from predefined binary features rather than engaging in open-ended inquiry, and the teacher's objective is fixed and known. The forms of student agency modeled here---selecting which feature to ask about---represent a specific and narrow kind of informational participation, not the broader notion of learner agency discussed in educational theory.
 
Second, the term ``universal task completion'' should not be interpreted as educational equity in the broader sense. What the simulation shows is that all students reach correct concept identification under certain interaction modes. It does not examine equity of access, participation, effort distribution, or broader distributional concerns. We use equalized task outcomes as a narrow, outcome-based indicator that is appropriate for the modeled setting but does not extend to the full meaning of equity in educational research. For instance, the model does not examine whether some students require more interaction steps, greater reliance on questioning, or bear a disproportionate communication burden---considerations that are central to a fuller account of educational fairness.
 
Third, the connection to sociocultural theory is best understood as \emph{motivational} rather than \emph{operational}. The model is inspired by the general principle that bidirectional interaction matters for learning, an idea that has deep roots in sociocultural and constructivist perspectives \citep{Vygotsky1978-um, Bruner1996-ys}. However, it does not implement scaffolding, zone of proximal development, or distributed cognition in any substantive sense. It implements Bayesian belief tracking and information-gain-based feature selection. Where we draw parallels to educational theory (as in Table~\ref{tab:construct_mapping}), these should be read as loose analogies that may guide future research, not as claims that the theory has been computationally formalized. Critically, the model does not represent social, material, or cultural context---dimensions that are central to sociocultural accounts of learning and teaching. This is a foundational modeling choice, not a secondary limitation: the framework deliberately isolates the informational structure of interaction by abstracting away everything else.
 
\subsection{Implications for AIED and Computational Modeling}
 
Despite these constraints, the findings suggest several directions worth investigating.
 
For the AIED community, the results illustrate---in a formal and transparent setting---how stopping rules based on incomplete information about heterogeneous learners can leave some learners unresolved. In teacher-initiated machine-teaching formulations, the learner receives examples selected by the teacher \citep{Zhu2015-js, Goldman1988-cw}. Our simulation shows that the baseline early-stopping teacher policies leave a subset of heterogeneous learners unresolved, whereas teacher-led continuation can recover complete coverage. These findings motivate examining how learner feedback, feature-selection policies and stopping rules jointly affect coverage and interaction cost in adaptive educational technologies. They do not establish that teacher-led instruction necessarily produces incomplete coverage. This is consistent with calls for cumulative theory development in instructional science \citep{Koedinger2012-vj} and with recent emphasis on interpretable and transparent approaches in AIED \citep{Balzan2025-uy}.
 
For the broader computational modeling community, the framework demonstrates that co-adaptive interaction can be studied in a formal, interpretable, and modular way. The teacher--student metaphor is widely used in machine learning---for instance, in knowledge distillation \citep{Hinton2015-la, Beyer2021-ql} and in multi-agent reinforcement learning \citep{Taylor2009-it, Silva2019-uz}---but these uses are typically unidirectional. Our framework shows that the tested interaction policies differ in coverage under their baseline stopping rules and in the interaction steps required to achieve complete identification. Teacher-led continuation also achieves complete coverage, while full co-adaptation reaches this outcome in fewer interaction steps on average at experience level 40. Whether similar principles apply in other multi-agent settings (e.g., human--AI collaboration, multi-agent coordination under asymmetric information) is an open question that the present work motivates but does not resolve.
 
We emphasize that these implications are \emph{hypotheses suggested by the simulation}, not conclusions established by it. Empirical validation in real educational settings is essential before any practical design recommendations can be drawn.

\section{Limitations and Future Directions}
\label{sec:limitations}
 
\subsection{Foundational Modeling Constraints}
 
\paragraph{Absence of social, material, and cultural context}
The model deliberately abstracts away the social, material, and cultural dimensions of education. This is a foundational modeling choice: the framework isolates the informational structure of teacher--student interaction by removing everything else. As a consequence, the results speak to the logic of co-adaptive information exchange under partial observability, but they do not address the situated, embodied, and culturally embedded nature of real teaching and learning \citep{Vygotsky1978-um, Gay2010-gt}.
 
\paragraph{Narrow operationalization of student heterogeneity}
Student differences are modeled as binary observability constraints over features. This captures one form of differential access to information, but it does not represent the cognitive, affective, motivational, or cultural diversity present in real classrooms. The model also assumes a fixed, known set of student types, whereas in practice, student differences are often unknown, emergent, and multidimensional.
 
\paragraph{Limited student agency}
Students in the model participate by selecting predefined features to ask about. This is a constrained form of query-based interaction, not the open-ended, expressive, or resistant forms of agency that characterize real learners. Students do not misunderstand, reinterpret, disengage, or pursue alternative goals. The model also does not include student-to-student interaction: while students in the same group hear each other's questions and the teacher's responses, there is no peer communication, negotiation, or co-construction.
 
\paragraph{Task simplicity}
The concept-learning task involves binary features, guaranteed learnability, and a finite concept space. These properties make the environment tractable and interpretable, but they are far removed from the open-ended, multi-level, and often ambiguous nature of real learning tasks.
 
\paragraph{Dependence on modeling assumptions}
The reported findings are conditional on the specific parameter values, concept-space structure, group compositions, stopping rules and probabilistic mechanisms used in the simulation. Completion times for unsuccessful policies are conditional on success and must be read alongside completion counts and all-episode trajectory coverage. Aggregate observability estimates can favour dominant types, as the within-group results show. Sequential learnability filtering is order-dependent and changes the task space across severity conditions, so it does not isolate observability at constant difficulty. We have examined robustness to observability severity (Table~\ref{tab:robustness}), but the framework has not been tested under all possible configurations. Future work should examine sensitivity to additional factors, including concept-space size, number of groups and alternative prior initializations; group-composition sensitivity is examined here.
 
\paragraph{Simulation uncertainty}
The primary comparison uses 100 independent runs with separate world and policy streams, preserving within-run dependence in uncertainty estimation. This characterises variability under the stated generator, not universality across possible environments. Perfect observed completion does not establish population certainty or equivalence; conservative boundary intervals are supplied.
 
\paragraph{No empirical validation}
The framework has not been validated against real classroom data. The findings are properties of the simulation and should be treated as formal hypotheses rather than empirically established facts.

\subsection{Future Directions}
 
\paragraph{Empirical validation}
The most important next step is to test whether the predictions generated by the framework correspond to patterns observable in real teacher--student interactions. Collaborations with educators could compare simulated interaction trajectories to classroom data, examining, for example, whether alternating teacher and student initiative improves learning outcomes for heterogeneous groups.
 
\paragraph{Richer interaction models}
Integrating more expressive forms of communication, for instance, via large language model (LLM)-based agents---could extend the framework to settings involving natural language, ambiguity, and richer forms of student feedback. Such extensions would move the framework closer to ecologically realistic teacher--student interaction while preserving the modular comparison structure.
 
\paragraph{Neurocomputational extensions}
Active inference \citep{Friston2015-qj, Buckley2017-rx} offers a principled alternative to the rational Bayesian framework used here. By formalizing cognition as minimization of variational free energy, active inference naturally unifies exploration and exploitation and can model phenomena such as belief inertia and social inhibition \citep{Maisto2023-vl, Ryan2001-vv}. Integrating active inference into the present framework could enhance both its biological plausibility and its capacity to model more nuanced forms of teacher--student dynamics.
 
\paragraph{Fairness-aware analysis}
The present study uses universal task completion as a narrow outcome-based indicator. Future work could examine richer fairness criteria, such as the distribution of interaction burden across student types, whether some students disproportionately depend on questioning, and whether co-adaptation introduces dependencies that affect some groups differently.
 
Despite these limitations, the framework's modularity and transparency make it amenable to principled extensions. Its primary value lies in providing a controlled, interpretable setting for generating formal hypotheses about the role of interaction structure in heterogeneous learning environments.

\section{Conclusion}
\label{sec:conclusion}
 
This paper introduced a Bayesian computational framework for studying co-adaptive teacher–student interaction under partial observability, evaluated in a stylized concept-learning task with heterogeneous student groups. Through systematic comparison of five interaction modes and three teacher-led controls across 100 independent matched simulation runs, the results show that learner-initiated queries resolve information gaps left by the baseline teachers’ stopping rules, while Bayesian belief tracking improves the efficiency of full co-adaptation as experience accumulates. At experience level 40, full co-adaptation achieves universal completion in all observed episodes and requires fewer interaction steps on average than active learning, turn-based interaction, and the three teacher-led controls (Table~\ref{tab:r2_controls}). These comparators also achieve complete observed coverage, so the advantage lies in reaching that outcome more efficiently. Together, these findings clarify the complementary roles of learner queries, teacher adaptation, and stopping rules within the modeled environment. 
 
The framework's primary contribution is methodological: it provides a formal, interpretable, and modular setting for studying how the structure of interaction affects learning outcomes in heterogeneous groups. It extends recent Bayesian models of adaptive teaching \citep{Chen2024-sy} to the one-to-many case with group-level belief tracking via Thompson sampling, and introduces a five-mode comparison structure that supports systematic comparison of explicit interaction mechanisms and stopping rules.
 
These findings are established within a deliberately simplified environment and do not constitute evidence about real classroom practice. However, they suggest formal hypotheses---such as the complementary roles of learner queries, teacher adaptation and continuation rules under heterogeneous information access---that may guide future empirical research. The most important next steps are empirical validation against real teacher--student interaction data, sensitivity analysis across a wider range of task environments, and extensions that incorporate richer forms of student agency and communication.

\section*{Acknowledgements}
We gratefully acknowledge Stefano Zingaro and Andrea Mata for their valuable feedback on the research direction, implications, limitations, and potential of this study.

\begin{samepage}
\section*{Funding}
F.B. was supported by Future AI Research (FAIR) PE01, SPOKE 8 on PERVASIVE AI, funded by the National Recovery and Resilience Plan \mbox{(NRRP)}.\par
\end{samepage}

\section*{CRediT authorship contribution statement}
\textbf{Francesco Balzan}: Conceptualization, Methodology, Investigation, Writing -- original draft, Supervision.\\
\textbf{Pedro P. Santos}: Software, Methodology, Investigation, Formal analysis, Visualization, Writing -- review \& editing.\\
\textbf{Maurizio Gabbrielli}: Writing -- review \& editing, Supervision.\\
\textbf{Mahault Albarracin}: Writing -- review \& editing, Investigation.\\
\textbf{Manuel Lopes}: Conceptualization, Methodology, Supervision, Writing -- review \& editing.

\section*{Declarations}

\begin{itemize}

\item \textbf{Conflict of Interest / Competing Interests}  
The authors declare no competing interests.

\item \textbf{Ethics Approval and Consent to Participate}  
Not applicable.

\item \textbf{Consent for Publication}  
Not applicable.

\item \textbf{Data Availability}  
The synthetic worlds and numerical outputs can be regenerated using Supplementary Code S1; the full mathematical specification and uncertainty derivations are supplied as Supplementary Methods S2.

\item \textbf{Materials Availability}  
Not applicable.

\item \textbf{Code Availability}  
Supplementary Code S1 is a self-contained Jupyter notebook providing executable code, configurations, environment requirements and instructions to regenerate the synthetic worlds and reproduce the reported numerical results and associated tables and figures.

\item \textbf{Declaration of generative AI and AI-assisted technologies in the manuscript preparation process} 
During the preparation and revision of this work, OpenAI's ChatGPT assisted translation, prose editing, code inspection, reproducibility checks and analysis scripting. The authors are responsible for verifying these outputs and for the final scientific content.

\end{itemize}

\appendix
\section{Experimental details}\label{app1}

\subsection{Experimental Scenario}
\begin{table}[H]
\centering
\caption{Experimental configuration. All parameter values are reported for reproducibility.}
\label{tab:exp_config}
\small
\resizebox{\textwidth}{!}{\begin{tabular}{ll}
\toprule
\textbf{Parameter} & \textbf{Value} \\
\midrule
\multicolumn{2}{l}{\textit{Concept space}} \\
Candidate concepts before filtering & 500 \\
Number of features $|\Phi|$ & 15 \\
Feature activation probabilities & [0.2, 0.3, 0.4, 0.5, 0.6, 0.7, 0.1, 0.4, \\
 & \phantom{[}0.4, 0.9, 0.8, 0.7, 0.5, 0.6, 0.2] \\
Actual concepts after filtering (across runs) & $M = 145.45$, $SD = 13.30$, range: 115--188 \\
Learnability enforced & Yes \\
\midrule
\multicolumn{2}{l}{\textit{Student heterogeneity}} \\
Number of student types $|\mathcal{T}|$ & 3 \\
Features unobservable per type & 5 (randomly selected per run) \\
\midrule
\multicolumn{2}{l}{\textit{Group structure}} \\
Number of groups & 3 \\
Students per group & 30 \\
Total students & 90 \\
Type distribution per group & 90\% dominant type, 5\%/5\% others \\
Student type assignment & Stochastic (per student, per run) \\
\midrule
\multicolumn{2}{l}{\textit{Teacher model}} \\
Feature selection (non-adaptive) & $\argmax_\phi$ information gain \\
Feature selection (adaptive) & $\argmax_\phi$ expected information gain \\
 & (weighted by sampled observability) \\
Belief tracking & $\text{Beta}(\alpha_{\phi,g}, \beta_{\phi,g})$ per feature per group \\
Prior initialization & $\alpha = 1, \beta = 1$ (uniform) \\
Exploration--exploitation & Thompson sampling \\
\midrule
\multicolumn{2}{l}{\textit{Simulation protocol}} \\
Independent primary runs & 100 matched across all eight policies \\
Target concepts taught per run & 41 \\
Maximum interaction steps per episode & 15 \\
Target concept sampling & Uniform random, with replacement \\
Master seed & 20261103; separate world/policy streams \\
Experience levels reported & 0, 20, 40 prior concepts taught \\
Confidence intervals & Whole-run 95\% bootstrap ($30{,}000$ resamples) \\
Sensitivity runs & 30 per condition; first 30 primary runs reused as reference \\
\bottomrule
\end{tabular}}
\end{table}

\begin{table}[H]
\centering
\caption{Information structure: what each agent knows and does not know.}
\label{tab:info_structure}
\small
\resizebox{\textwidth}{!}{\begin{tabular}{p{1.8cm} p{5.5cm} p{5.5cm}}
\toprule
\textbf{Agent} & \textbf{Known} & \textbf{Unknown} \\
\midrule
Teacher & Target concept and its feature values; group identity; features previously suggested; whether the selected respondent observed each suggested feature & Type of each student; which features each student can observe; how many students have correctly identified the target \\
\addlinespace
Student & Feature values they can observe (given their type); all suggestions and answers exchanged within the group & Features they cannot observe; the target concept (initially); types and observations of other students \\
\bottomrule
\end{tabular}}
\end{table}

\subsection{Experimental Modes}

\begin{table}[H]
\centering
\caption{Interaction Modes.}
\label{tab:strategy_comparison}
\footnotesize
\setlength{\tabcolsep}{3pt}
\begin{tabular}{lccccp{2.5cm}}
\toprule
\textbf{Mode} & \textbf{Teacher} & \textbf{Students} & \textbf{Adapts} & \textbf{Belief} & \textbf{Analogous} \\
               & \textbf{initiates} & \textbf{query} & \textbf{to groups} & \textbf{tracking} & \textbf{setting} \\
\midrule
Active learning     & No  & Yes & No  & No  & Student-led inquiry \\
Active teaching     & Yes & No  & No  & No  & Uniform lecture \\
Turn-based          & Yes $\to$ No & Yes & No  & No  & Lecture + Q\&A \\
Adaptive teaching   & Yes & No  & Yes & Yes & Differentiated instruction \\
Full co-adaptation  & Yes $\to$ No & Yes & Yes & Yes & Diff.\ instruction + inquiry \\
\bottomrule
\end{tabular}
\end{table}

\section{Additional results}\label{app:r2analyses}
\subsection{Teacher-led controls}\label{app:r2controls}
After the baseline active or adaptive teacher stops, its continuation control reveals unannounced features in ascending index order. Exhaustive teaching uses that order from the beginning. All policies share the same worlds, targets and maximum of 15 actual interaction steps, with no hidden learner beliefs supplied to teacher selection. Learnability filtering makes exhaustive coverage sufficient for every type.
\input{r2_control_table.tex}

\subsection{Outcomes within groups}\label{app:r2subgroups}

\begin{table}[!htbp]\centering\footnotesize
\caption{Group-level results at experience 40, on 100 matched runs. Dominant types are fixed by group position, with stochastic 90\%/5\%/5\% student assignments. Interaction steps to completion exclude unsuccessful episodes.}
\label{tab:group_results}
\resizebox{\textwidth}{!}{\begin{tabular}{llrrr}\toprule
Group & Mode & Final \% correct (SD) & Completed/total & Interaction steps (SD) \\\midrule
Group 1 & Active learning & $100.00 \pm 0.00$ & 100/100 & $10.18 \pm 1.66$ \\
Group 1 & Active teaching & $86.33 \pm 25.41$ & 22/100 & $9.23 \pm 2.50$ \\
Group 1 & Turn-based & $100.00 \pm 0.00$ & 100/100 & $9.80 \pm 2.10$ \\
Group 1 & Adaptive teaching & $92.20 \pm 5.25$ & 11/100 & $5.91 \pm 1.24$ \\
Group 1 & Full co-adaptation & $100.00 \pm 0.00$ & 100/100 & $9.21 \pm 2.16$ \\
\midrule
Group 2 & Active learning & $100.00 \pm 0.00$ & 100/100 & $10.11 \pm 1.78$ \\
Group 2 & Active teaching & $75.90 \pm 37.24$ & 26/100 & $8.23 \pm 2.12$ \\
Group 2 & Turn-based & $100.00 \pm 0.00$ & 100/100 & $9.66 \pm 2.02$ \\
Group 2 & Adaptive teaching & $91.20 \pm 6.14$ & 13/100 & $6.15 \pm 2.18$ \\
Group 2 & Full co-adaptation & $100.00 \pm 0.00$ & 100/100 & $9.20 \pm 2.09$ \\
\midrule
Group 3 & Active learning & $100.00 \pm 0.00$ & 100/100 & $9.91 \pm 1.71$ \\
Group 3 & Active teaching & $79.93 \pm 33.46$ & 24/100 & $8.33 \pm 2.03$ \\
Group 3 & Turn-based & $100.00 \pm 0.00$ & 100/100 & $9.70 \pm 2.02$ \\
Group 3 & Adaptive teaching & $89.47 \pm 16.72$ & 16/100 & $6.19 \pm 2.27$ \\
Group 3 & Full co-adaptation & $100.00 \pm 0.00$ & 100/100 & $8.96 \pm 2.13$ \\
\bottomrule\end{tabular}}\end{table}

Table~\ref{tab:r2_subgroups} retains all nine group/type cells; pooled role estimates are in the main text.
\input{r2_types_compact.tex}

\subsection{Strategy comparisons and sensitivity analyses}\label{app:r2contrasts}\label{app:r2counts}
\begin{figure}[H]\centering\includegraphics[width=\textwidth]{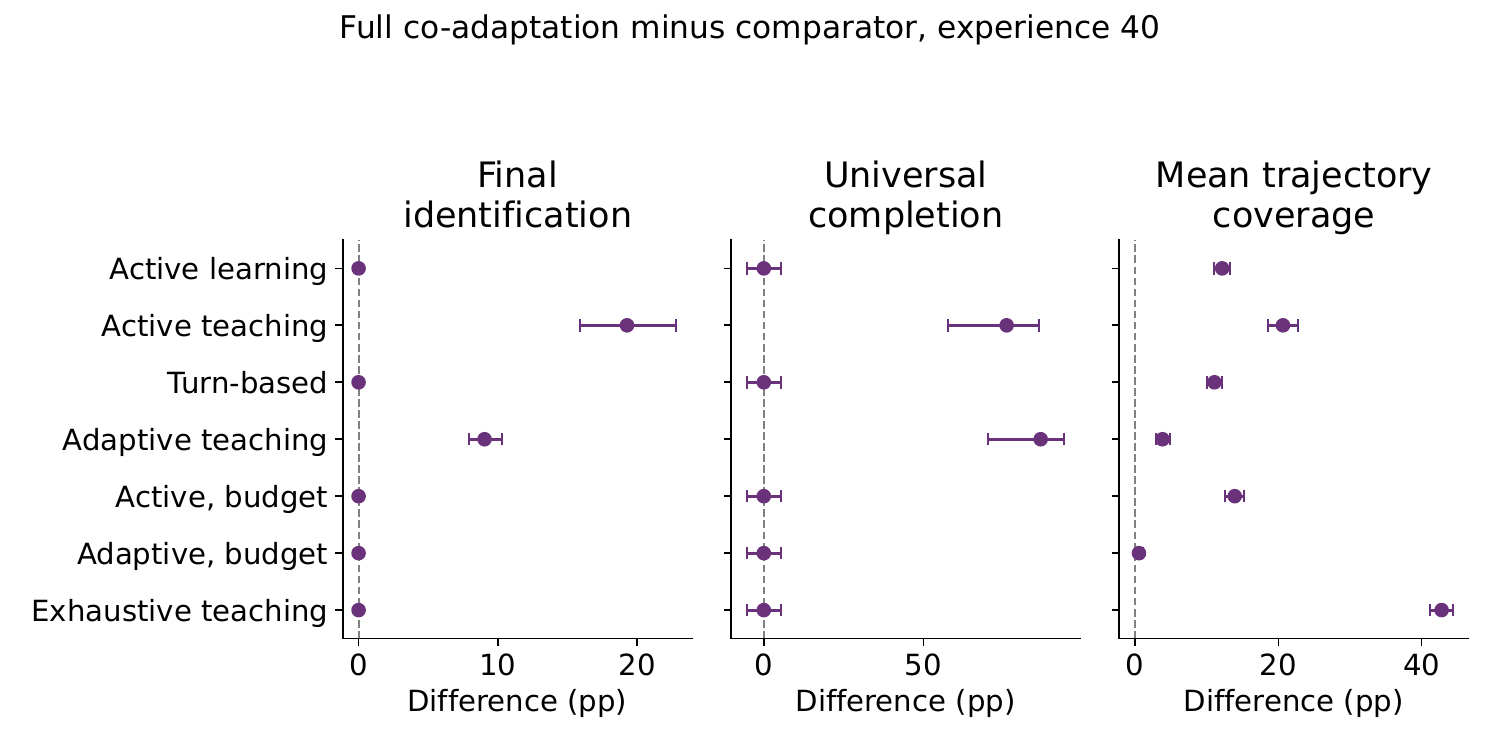}
\caption{Full co-adaptation minus comparator at experience 40, with pointwise 95\% intervals. Identification and trajectory coverage use paired whole-run resampling; completion uses conservative group-position binomial bounds. These are not simultaneous or equivalence intervals.}\label{fig:r2_differences}\end{figure}

Figure~\ref{fig:r2_differences} displays direct policy contrasts. Table~\ref{tab:r2_counts} characterises the retained task spaces. These diagnostics do not fully equate difficulty across observability conditions; the complete sensitivity results and exhaustive-control costs are in Supplementary Code S1.

\begin{table}[!htbp]\centering\footnotesize
\caption{Observability-severity sensitivity at experience 40: 30 runs, hence 90 group-episodes, per condition. The five core strategies are shown; all eight policies are included in the executable outputs. The five-disabled reference reuses the first 30 primary runs. Times are conditional on completion.}
\label{tab:robustness}
\resizebox{\textwidth}{!}{\begin{tabular}{clrrr}\toprule
Disabled & Mode & Final \% correct (SD) & Completed/total & Interaction steps (SD) \\\midrule
3/15 & Active learning & $100.00 \pm 0.00$ & 90/90 & $10.24 \pm 1.36$ \\
3/15 & Active teaching & $89.78 \pm 21.81$ & 31/90 & $8.23 \pm 2.11$ \\
3/15 & Turn-based & $100.00 \pm 0.00$ & 90/90 & $9.37 \pm 2.08$ \\
3/15 & Adaptive teaching & $91.81 \pm 14.80$ & 22/90 & $7.18 \pm 1.85$ \\
3/15 & Full co-adaptation & $100.00 \pm 0.00$ & 90/90 & $9.06 \pm 2.14$ \\
\midrule
5/15 & Active learning & $100.00 \pm 0.00$ & 90/90 & $9.73 \pm 1.50$ \\
5/15 & Active teaching & $82.11 \pm 31.44$ & 22/90 & $8.86 \pm 1.84$ \\
5/15 & Turn-based & $100.00 \pm 0.00$ & 90/90 & $9.67 \pm 1.72$ \\
5/15 & Adaptive teaching & $89.41 \pm 14.68$ & 10/90 & $5.80 \pm 1.25$ \\
5/15 & Full co-adaptation & $100.00 \pm 0.00$ & 90/90 & $8.90 \pm 1.84$ \\
\midrule
7/15 & Active learning & $100.00 \pm 0.00$ & 90/90 & $9.09 \pm 1.31$ \\
7/15 & Active teaching & $87.04 \pm 25.50$ & 28/90 & $9.39 \pm 2.09$ \\
7/15 & Turn-based & $100.00 \pm 0.00$ & 90/90 & $9.59 \pm 1.47$ \\
7/15 & Adaptive teaching & $89.74 \pm 11.26$ & 4/90 & $7.50 \pm 2.69$ \\
7/15 & Full co-adaptation & $100.00 \pm 0.00$ & 90/90 & $8.60 \pm 1.53$ \\
\bottomrule\end{tabular}}\end{table}

\input{r2_count_table.tex}

\paragraph{Uncertainty}\label{app:statistics}
Intervals resample whole runs and recompute student-type ratios. Completion bounds use group-position binomial intervals with a union-bound correction; Supplementary Methods S2, Section~2 gives the derivation \citep{ClopperPearson1934, Morris2019}.

\end{document}

%% file: r2_primary_table.tex
\begin{table}[!htbp]\centering\footnotesize
\caption{Quantitative results summary across experience levels. Final identification and completion include all 300 group-episodes. Interaction steps to completion are conditional on success; successful/total counts are explicit. Mean coverage averages identification across all 15 plotted steps, carrying terminal states forward. SDs describe variation across group-episodes.}
\label{tab:results_summary}
\resizebox{\textwidth}{!}{\begin{tabular}{clrrrr}\toprule
Exp. & Interaction mode & Final \% correct (SD) & Completed/total & Interaction steps (SD) & Mean coverage \% \\\midrule
0 & Active learning & $100.00 \pm 0.00$ & 300/300 & $9.84 \pm 1.54$ & 54.39 \\
0 & Active teaching & $85.12 \pm 27.93$ & 84/300 & $8.50 \pm 2.30$ & 49.92 \\
0 & Turn-based & $100.00 \pm 0.00$ & 300/300 & $9.35 \pm 1.96$ & 57.47 \\
0 & Adaptive teaching & $81.47 \pm 32.98$ & 102/300 & $10.18 \pm 2.31$ & 39.23 \\
0 & Full co-adaptation & $100.00 \pm 0.00$ & 300/300 & $10.03 \pm 1.84$ & 51.55 \\
\midrule
20 & Active learning & $100.00 \pm 0.00$ & 300/300 & $9.98 \pm 1.60$ & 54.44 \\
20 & Active teaching & $82.40 \pm 30.74$ & 75/300 & $8.32 \pm 2.23$ & 48.52 \\
20 & Turn-based & $100.00 \pm 0.00$ & 300/300 & $9.44 \pm 2.10$ & 56.54 \\
20 & Adaptive teaching & $90.76 \pm 14.43$ & 51/300 & $6.45 \pm 2.58$ & 62.40 \\
20 & Full co-adaptation & $100.00 \pm 0.00$ & 300/300 & $8.83 \pm 2.13$ & 66.69 \\
\midrule
40 & Active learning & $100.00 \pm 0.00$ & 300/300 & $10.07 \pm 1.72$ & 53.92 \\
40 & Active teaching & $80.72 \pm 32.70$ & 72/300 & $8.57 \pm 2.26$ & 45.42 \\
40 & Turn-based & $100.00 \pm 0.00$ & 300/300 & $9.72 \pm 2.05$ & 55.00 \\
40 & Adaptive teaching & $90.96 \pm 10.78$ & 40/300 & $6.10 \pm 2.01$ & 62.25 \\
40 & Full co-adaptation & $100.00 \pm 0.00$ & 300/300 & $9.12 \pm 2.13$ & 66.06 \\
\bottomrule\end{tabular}}\end{table}

%% file: r2_control_table.tex
\begin{table}[H]\centering\footnotesize
\caption{Teacher-led controls at experience 40: 100 matched runs, 300 group-episodes each. All reach 100\% identification. Coverage averages identification across 15 plotted steps. The final column is full co-adaptation minus control, with paired-run 95\% intervals over the full common episode set. Other checkpoints and contrasts are in Supplementary Code S1.}
\label{tab:r2_controls}\label{tab:r2_time}
\resizebox{\textwidth}{!}{\begin{tabular}{lrrrr}\toprule
Control & Completed & Interaction steps (SD) & Coverage \% & $\Delta$ interaction steps [95\% CI] \\\midrule
Active + continuation & 300/300 & 11.78 (2.76) & 52.16 & -2.65 [-2.94, -2.37] \\
Adaptive + continuation & 300/300 & 11.88 (3.03) & 65.54 & -2.75 [-3.02, -2.49] \\
Exhaustive teaching & 300/300 & 13.69 (1.32) & 23.26 & -4.56 [-4.85, -4.27] \\
\bottomrule\end{tabular}}\end{table}

%% file: r2_types_compact.tex
\begin{table}[H]\centering\footnotesize
\caption{Within-group type outcomes at experience 40; Type $g$ is dominant in Group $g$. $N$ counts students and $n$ identifications; intervals resample runs. Active learning, turn-based, full co-adaptation and all three controls have $n=N$ in every row. AT denotes active teaching and AdT adaptive teaching. Boundary empirical intervals are not population guarantees. Full eight-strategy results are in Supplementary Code S1.}
\label{tab:r2_subgroups}
\resizebox{\textwidth}{!}{\begin{tabular}{llrrrrr}\toprule
Group & Type & $N$ & $n_{\rm AT}$ & AT \% [95\% CI] & $n_{\rm AdT}$ & AdT \% [95\% CI] \\\midrule
1 & 1 & 2732 & 2516 & 92.09 [86.28, 97.03] & 2732 & 100.00 [100.00, 100.00] \\
1 & 2 & 148 & 36 & 24.32 [14.79, 35.54] & 18 & 12.16 [4.32, 21.68] \\
1 & 3 & 120 & 38 & 31.67 [20.16, 43.97] & 16 & 13.33 [4.88, 23.21] \\
\midrule
2 & 1 & 167 & 45 & 26.95 [16.95, 38.04] & 16 & 9.58 [3.61, 16.76] \\
2 & 2 & 2690 & 2173 & 80.78 [72.79, 88.17] & 2690 & 100.00 [100.00, 100.00] \\
2 & 3 & 143 & 59 & 41.26 [29.53, 53.55] & 30 & 20.98 [11.83, 31.08] \\
\midrule
3 & 1 & 132 & 53 & 40.15 [27.27, 52.74] & 20 & 15.15 [5.93, 25.74] \\
3 & 2 & 145 & 46 & 31.72 [20.57, 43.84] & 22 & 15.17 [7.35, 24.14] \\
3 & 3 & 2723 & 2299 & 84.43 [77.16, 91.23] & 2642 & 97.03 [93.16, 100.00] \\
\bottomrule\end{tabular}}\end{table}

%% file: r2_count_table.tex
\begin{table}[H]\centering\small
\caption{Actual retained concepts and initial uncertainty in the simulations. Five-disabled/high-homogeneity sensitivity uses the first 30 primary worlds. Conditions with different observability need not have equal task difficulty.}
\label{tab:r2_counts}
\resizebox{\linewidth}{!}{\begin{tabular}{lrrrr}\toprule
Condition & Worlds & Concepts: mean (SD) & Range & Mean $\log_2 N_c$ \\\midrule
Primary & 100 & 145.45 (13.30) & 115--188 & 7.18 \\
5 disabled / high homogeneity & 30 & 143.60 (13.45) & 115--170 & 7.16 \\
3 disabled & 30 & 267.70 (16.21) & 226--301 & 8.06 \\
7 disabled & 30 & 56.63 (5.44) & 48--70 & 5.82 \\
Medium homogeneity & 30 & 146.43 (15.54) & 118--178 & 7.19 \\
Uniform mixture & 30 & 147.07 (14.52) & 124--182 & 7.19 \\
\bottomrule\end{tabular}}\end{table}